\documentclass[12pt,psfig]{article}
\usepackage{graphicx}

\begin{document}

\begin{center} 
{\bf  The  ``ab initio" approach to                                                       
 the nuclear equation of state: review and discussion 
 }              
\end{center}
\vspace{0.1cm} 
\begin{center} 
 Francesca Sammarruca \\ 
\vspace{0.2cm} 
 Physics Department, University of Idaho, Moscow, ID 83844, U.S.A   
\end{center} 
\begin{abstract}
We review the main components of our microscopic model of nuclear matter, which
we have recently extended to incorporate isospin-asymmetry.               
Some frequently discussed issues concerning nuclear many-body approaches are revisited and critically analysed. 
\\ \\ 
PACS number(s): 21.65.+f,21.30.Fe 
\end{abstract}

\section{Introduction} 
                                                                     
Issues related to 
the nuclear equation of state (EoS), in particular the one                                          
describing isospin-asymmetric nuclear matter, 
are gathering increased interest within the nuclear physics
community.                                                              
Heavy-ion (HI) collision observables which are sensitive to the symmetry energy are being identified
and used to constrain this fundamentally important quantity \cite{HI1,HI2,HI3,HI4,HI5}. 
The neutron skin in neutron-rich nuclei is also related to some features of the EoS, 
specifically the difference between the pressure in neutron matter and in 
symmetric matter, that is, the pressure gradient that pushes neutrons
outwards to form the skin. 
Therefore, empirical information
on the structure of these nuclei can help constrain the shape of the symmetry energy.
Parity-violating electron scattering experiments \cite{skin} appear to be the most promising way to obtain 
information on neutron densities in the near future. 
Vice versa, independent reliable constraints on the density dependence of the symmetry energy 
would facilitate predictions of neutron skins. 

In this paper, 
we like to suggest that ``{\it ab initio}" calculations are the best way to complement such rich
and intense experimental and phenomelogical efforts. ``{\it Ab initio}" means that the starting point are   
realistic free-space nucleon-nucleon (NN) interactions (potentially complemented by many-body forces)    
which are then applied in the nuclear many-body system.        
Phenomenological interactions, such as Skyrme forces (see Ref.~\cite{Sk} and references therein), or various parametrizations of relativistic mean field models (see Ref.~\cite{RMF} and references therein), may not be able to provide sufficient 
physical insight. Of course,                              
phenomenological models are practical and very useful as a testing tool, for instance, 
when setting empirical boundaries to the EoS through analyses of HI collisions.   
But, ultimately, comparison between constraints and theoretical predictions is important for an
understanding on a more fundamental level. Such comparison will help identify strengths and 
weaknesses of the theoretical models and at the same time provide insight into the physical relevance of the 
``observable" under consideration. 

In the next sections, we review the main aspects of our microscopic approach, starting from the 
two-body input and proceeding into our nuclear matter calculations, which are based on the 
Dirac-Brueckner-Hartree-Fock (DBHF) method. 
At each step, 
we will discuss and motivate our choices.

\section{The  ``ab initio" approach} 
\subsection{The two-body framework} 
Our present knowledge of the nuclear force is the result of decades of
struggle \cite{Mac89}. After the development of QCD and the understanding of its symmetries,  
chiral effective theories \cite{chi} became popular as a way to respect the       
symmetries of QCD while keeping the degrees of freedom (nucleons and pions) typical of low-energy nuclear physics. However, 
chiral perturbation theory (ChPT)
has definite limitations as far as the range of allowed momenta is concerned. 
For the purpose of applications in dense matter, where higher and higher momenta become involved     
with increasing Fermi momentum, ChPT is inappropriate. 
A relativistic, meson-theoretic model is the better choice.                    

The one-boson-exchange (OBE) model has proven very successful in describing NN data in free space 
and has a good theoretical foundation. 
Among the many available OBE potentials (some being part of the ``high-precision generation"              
\cite{pot1,pot2,pot3}), 
we seek a momentum-space potential developed within a relativistic scattering equation, such as the 
one obtained through the Thompson three-dimensional reduction of the Bethe-Salpeter equation.
Furthermore, we require a potential that uses 
the pseudovector coupling for the interaction of nucleons with pseudoscalar mesons. 
With this in mind, 
as well as the requirement of a good description of NN data, 
Bonn B \cite{Mac89} has been our standard choice. As is well known, the NN potential model dependence
of nuclear matter predictions is not negligible. The saturation points obtained with different NN potentials
move along the famous ``Coester band" depending on the strength of the tensor force, with the weakest tensor
force corresponding to the largest attraction. For the same reason (that is, the role of the tensor force in  
nuclear matter), 
the potential model dependence is strongly reduced in pure (or nearly pure) neutron matter, due to the  
absence of isospin-zero partial waves. 

Already when QCD (and its symmetries) were unknown, it was observed that the contribution from the
nucleon-antinucleon pair diagram, Fig.~1, is unreasonably large when the pseudoscalar (ps) coupling is used, 
leading to very large pion-nucleon scattering lengths \cite{GB79}.                                            
We recall that the Lagrangian density for pseudoscalar coupling of the nucleon field ($\psi$) with the  pseudoscalar meson
field ($\phi$) is 
\begin{equation}
{\cal L}_{ps} = -ig_{ps}\bar {\psi} \gamma _5 \psi \phi. 
\end{equation} 
On the other hand, the same contribution (Fig.~1) 
is heavily suppressed by the pseudovector (pv) coupling (a mechanism which
became known as ``pair suppression"). The reason for the suppression is the presence of the 
covariant derivative (that is, a four-momentum dependence)                                               
at the pseudovector vertex,                                                  
\begin{equation}
{\cal L}_{pv} = \frac{f_{ps}}{m_{ps}}{\bar \psi}  \gamma _5 \gamma^{\mu}\psi \partial_{\mu} \phi, 
\end{equation} 
which reduces the contribution of the diagram. Because $\partial _{\mu}$ is equivalent to the 
momentum $q_{\mu}$ (in momentum space), the equation above explains the small values of the pion-nucleon
scattering length at threshold \cite{GB79}. 
Considerations based on chiral symmetry \cite{GB79} can further motivate 
the choice of the pseudovector coupling.                          
We will come back to this point in the next section. 

The most important aspect of the ``{\it ab initio}" approach is that the only free parameters of the
model (namely, the parameters of the NN potential)                                               
are determined by the fit to the free-space data and never readjusted in the medium. In other
words, the model parameters are tightly constrained and the calculation in the medium is 
parameter free. 
The presence of free parameters in the medium would generate effects and sensitivities which can be very large and hard to
control. 

\begin{figure}
\begin{center}
\vspace*{-4.0cm}
\hspace*{-2.0cm}
\scalebox{0.8}{\includegraphics{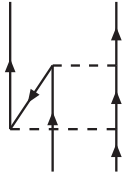}}
\vspace*{-17.0cm}
\caption{Contribution to the NN interaction from virtual pair excitation.                   
Upward- and downward-pointing arrows represent nucleons and antinucleons, respectively.
Dashed lines denote mesons.                            
} 
\label{one}
\end{center}
\end{figure}
\subsection{The many-body framework: Brueckner theory, three-body forces, and relativity} 

Excellent reviews of Brueckner theory have been written which we can refer the reader to 
(see \cite{Mac89} and references therein). 
Here, we begin by defining the contributions that are retained in our calculation. Those are 
the lowest order contribution to the Brueckner series (two-hole lines) and the corresponding
exchange diagram. 
With the G-matrix as the effective interaction, this amounts to including particle-particle
(that is, short-range) correlations, which are absolutely essential to even approach a realistic
description of nuclear matter properties. 
Three-nucleon correlations have been shown to be small if the continuous choice is adopted 
for the single-particle potential \cite{Baldo98}. 

The issue of three-nucleon forces (3NF), of course, remains to be discussed. 
In Fig.~2 we show a 3NF originating from virtual excitation of a nucleon-antinucleon pair, 
known as the ``Z-diagram". Notice that the observations from the previous section ensures that the corresponding diagram
at the two-body level, Fig.~1, is small with pv coupling. 
At this point, it is useful 
to recall the main feature of the Dirac-Brueckner-Hartree-Fock 
(DBHF) method, as that turns out to be closely related to 
the 3NF depicted in Fig.~2. In the DBHF approach, one describes the positive energy solutions
of the Dirac equation in the medium as 
\begin{equation}
u^*(p,\lambda) = \left (\frac{E^*_p+m^*}{2m^*}\right )^{1/2}
\left( \begin{array}{c}
 {\bf 1} \\
\frac{\sigma \cdot \vec {p}}{E^*_p+m^*} 
\end{array} 
\right) \;
\chi_{\lambda},
\end{equation}
where the effective mass is given by $m^* = m+U_S$, with $U_S$ an attractive scalar potential.
It turns out that both the description of single-nucleon propagation via Eq.~(3) and the evaluation of the 
Z-diagram, Fig.~2, generate a repulsive effect on the energy/particle in symmetric nuclear matter which depends on the density approximately
as 
\begin{equation}
\Delta E \propto  \left (\frac{\rho}{\rho_0}\right )^{8/3}, 
\end {equation}
and provides the saturating mechanism missing from conventional Brueckner calculations. 
Brown showed that the bulk of this effect can be obtained as a lowest order (in $p^2/m$) relativistic correction
to the single-particle propagation \cite{GB87}. 

\begin{figure}
\begin{center}
\vspace*{-4.0cm}
\hspace*{-2.0cm}
\scalebox{0.8}{\includegraphics{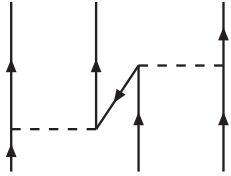}}
\vspace*{-17.0cm}
\caption{Three-body force due to virtual pair excitation. Conventions as in the previous figure.
} 
\label{two}
\end{center}
\end{figure}

The approximate equivalence of the effective-mass spinor description and the contribution from the Z-diagram 
has a simple intuitive explanation in the observation 
that Eq.~(3), like any other solution of the Dirac equation,
can be written as a combination of positive and negative energy solutions. On the other hand, the ``nucleon" in the 
middle of the Z-diagram, Fig.~2, is precisely a superposition of positive and negative energy states. 
In summary, the DBHF method effectively takes into account a particular class of 
3NF, which are crucial for nuclear matter saturation. 
Notice that the effective mass {\it ansatz} just outlined is extended to 
deal with protons and neutrons in different concentrations in the case of asymmetric matter       
\cite{AS03}.

Other, more popular, three-body forces need to be examined as well. 
Figure~3 shows the 3NF that is included in essentially all 3NF models, regardless
other components; it is the Fujita-Miyazawa 3NF \cite{FM}.                                          
With the addition of contributions from $\pi N$ S-waves, one ends up with the 
well-known Tucson-Melbourne 3NF \cite{TM}. The microscopic 3NF of Ref.~\cite{Catania} include 
contributions from excitations of the Roper resonance (P$_{11}$ isobar) as well. 
Now, if diagrams such as the one shown on the left-hand side of  Fig.~3 are included, 
consistency requires that medium modifications at the corresponding two-body level are also included, 
that is, the diagram on the right-hand side of Fig.~3 should be present and properly medium modified. 
Large cancellations then take place, a fact that was brought up a long time ago \cite{DMF} but perhaps not fully 
appreciated. When the two-body sector is handled via OBE diagrams, the two-pion exchange is 
effectively incorporated through the $\sigma$ ``meson", which 
certainly cannot generate the (large) medium effects (dispersion and Pauli blocking on $\Delta$
intermediate states) required by the consistency arguments above. 

Finally, one may wonder whether other 3N forces are being overlooked that might change the above scenario
in a significant way. 
Although a definite answer to the question of which 3NF should be included                           
can only come from chiral perturbation theory (at each 
order), in meson theory one can find guidance from considerations of range. Given that there are 
form factors at the vertices, inclusion of 3NF diagrams such as the one shown 
in Fig.~3, with both $\pi$ and $\rho$, should be able to account for the major 
3NF contribution. There are, of course, 
other 3NF, such as the three-pion rings included in the Illinois 3NF \cite{Ill3NF}. Those have been found to be
helpful in subtleties of the spin dependence when calculating spectra of light nuclei,            
but it seems unlikely that they would have a major impact on the properties of nuclear matter. 

\begin{figure}
\begin{center}
\vspace*{-4.0cm}
\hspace*{-2.0cm}
\scalebox{0.8}{\includegraphics{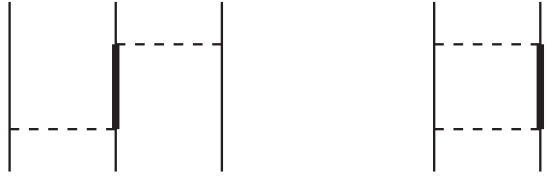}}
\vspace*{-17.0cm}
\caption{Left: three-body force arising from $\Delta$-isobar excitation (thick line). 
Right: two-meson exchange contribution to the NN interaction involving 
$\Delta$-isobar excitation.
} 
\label{three}
\end{center}
\end{figure}
In summary, our calculation does not include 3NF, except for those from virtual pair                              
excitations, which are accounted for indirectly. Collecting all the considerations above, 
we conclude that the DBHF method may be a reliable, yet practical many-body framework, and, possibly,   
more internally consistent than other microscopic approaches which include explicit 3NF.

\section{Conclusions}                                
It is the purpose of this note to underline the importance of ``{\it ab initio}" calculations
of the EoS to complement on-going experimental efforts. 
Several microscopic models are available which include either microscopic \cite{Catania} or 
phenomenological \cite{APR} 3NF or are based on the DBHF scheme \cite{Fuchs,Sam08} (limiting ourselves
to models that have recently been concerned with asymmetric matter).
We have re-examined 
some issues frequently encountered in the literature concerning ``popular" 3NF. Such 
3NF are absent from our (DBHF) calculation for reasons of consistency with the two-body sector,
whereas those originating from virtual nucleon-antinucleon excitation are inherent to the 
DBHF scheme. 
It is in fact          
remarkable that a relativistic effect can be shown to be essentially equivalent to a many-body force. 

In conclusion, 
microscopic calculations of the EoS and stringent constraints from EoS-sensitive 
observables can reveal information about the nature of the underlying nuclear force 
and its behavior in the medium.                                                
With the wealth of experiments/analyses presently going on or planned for the near future, and coherent effort
from experiment and theory, the prospects of a significant improvement in our knowledge 
of the equation of state, particularly its isospin asymmetries, 
are very good.

\section*{Acknowledgments}
Support from the U.S. Department of Energy under Grant No. DE-FG02-03ER41270 is 
acknowledged.                                                                           

\end{document}